\documentclass[aps,prb,groupedaddress,onecolumn,longbibliography,reprint]{revtex4-2}

\usepackage{graphicx}
\usepackage{amsmath,amsfonts,amssymb,mathtools}
\usepackage{color}
\usepackage{subfigure}
\usepackage{physics}
\usepackage{hyperref}
\usepackage{url}
\usepackage{booktabs}
\usepackage[dvipsnames]{xcolor}

\definecolor{emerald}{rgb}{0.31, 0.78, 0.47}
\definecolor{blue(ncs)}{rgb}{0.0, 0.53, 0.74}

\hypersetup{
     colorlinks=true,
     linkcolor=blue(ncs),
     filecolor=blue,
     citecolor=emerald,      
     urlcolor =blue(ncs),
}
	    
\definecolor{lightgray}{rgb}{0.9,0.9,0.9}	    
\definecolor{green}{rgb}{0,0.5,0}
\definecolor{red}{rgb}{1,0,0}
\definecolor{blue}{rgb}{0,0,0.5}

\long\def\symbolfootnote[#1]#2{\begingroup%
\def\thefootnote{\fnsymbol{footnote}}\footnotetext[#1]{#2}\footnotemark[#1]\endgroup}

\DeclareMathAlphabet{\pazocal}{OMS}{zplm}{m}{n}

\newcommand{\bo}[1]{\boldsymbol{#1}}

\newcommand{\D}{\mathrm{d}}
\newcommand{\f}[2]{\frac{#1}{#2}}

\begin{document}

\title{Nonlocal electrodynamics and the penetration depth of superconducting Sr$_2$RuO$_4$}

\author{Henrik S.~R{\o}ising}
  \affiliation{Niels Bohr Institute, University of Copenhagen, DK-2200 Copenhagen, Denmark}

\author{Andreas Kreisel}
\affiliation{Niels Bohr Institute, University of Copenhagen, DK-2200 Copenhagen, Denmark}   

\author{Brian M.~Andersen}
\affiliation{Niels Bohr Institute, University of Copenhagen, DK-2200 Copenhagen, Denmark}  

\date{August 8, 2024}

\begin{abstract}
The thermal quasiparticles in a clean type-II superconductor with line nodes give rise to a quadratic low-temperature change of the penetration depth, $\Delta \lambda \sim T^2$, as first shown by Kosztin and Leggett [I.~Kosztin and A.~J.~Leggett, Phys.~Rev.~Lett.~\textbf{79}, 135 (1997)]. Here, we generalize this result to multiple nodes and compare it to numerically exact evaluations of the temperature-dependent penetration depth in Sr$_2$RuO$_4$ using a high-precision tight-binding model. We compare the calculations to recent penetration depth measurements in high purity single crystals of Sr$_2$RuO$_4$ [J.~F.~Landaeta et al.,~\href{https://ui.adsabs.harvard.edu/abs/2023arXiv231205129L}{arXiv:2312.05129}]. When assuming the order parameter to have $\mathrm{B}_{1\mathrm{g}}$ symmetry, we find that both a simple $d_{x^2-y^2}$-wave and complicated gap structures with contributions from higher harmonics and accidental nodes can accommodate the experimental data.
\end{abstract}

\maketitle

%
\section{Introduction}
%
Obtaining a detailed understanding of the superconducting state in Sr$_2$RuO$_4$ remains an important outstanding problem~\cite{MaenoEA24, PustogowEA19, MackenzieEA17}. Experimentally, the quest is complicated due to challenging material properties of Sr$_2$RuO$_4$ and the low energy scale of the superconducting phase. Theoretically, Sr$_2$RuO$_4$ provides an important testbed for modeling of unconventional superconductivity. More specifically, the well-characterized electronic properties of the normal state offers a rather unique opportunity to test various electronic fluctuation-based mechanisms for pairing against detailed investigations of the superconducting gap structure~\cite{ScaffidiEA14,ZhangEA18,RoisingEA19,Ramires2019,LindquistKee19,WangZhang,GingrasEA19,RomerEA19,SuhEA19,Kivelson2020,RomerEA20,ASR2,Clepkens2021,WangEA22,ATR1,Roig2022,Roising2022,Kaser2022,Gingras2022,JerzembeckEA21}.

The magnetic penetration depth, $\lambda$, can elucidate nodal features of the gap, as revealed through its dependency on temperature and disorder~\cite{Prozorov_2006,Prozorov_2011,CARRINGTON2011502,Hirschfeld2016}. Fully-gapped superconductors exhibit exponential temperature dependence at low temperatures, whereas nodal gaps display power law dependence. In the case of a simple d-wave superconductor, deviations linear in the temperature $T$ are expected and a canonical interpretation of a $T^2$ dependence in d-wave superconductors is impurities.
In Sr$_2$RuO$_4$, measurements of the change in penetration depth, $\Delta \lambda(T) \equiv \lambda(T) - \lambda(0)$, via a tunnel diode oscillator method also yielded $\Delta \lambda(T) \sim T^2$, signalling the existence of nodal quasiparticle excitations~\cite{BonaldeEA00}. Renewed measurements of $\Delta \lambda(T)$ in Sr$_2$RuO$_4$ high-purity spherical single crystals using  scanning SQUID microscopy~\cite{MuellerEA23} and ac-susceptibility measurements~\cite{LandaetaEA23} have confirmed this behavior, excluded interpretation in terms of impurity effects and highlighted the importance of nonlocal Meissner screening. Indeed, as initially demonstrated by Kosztin and Leggett, nonlocal effects change the low temperature dependence of the penetration depth, e.g.,~from linear to quadratic in $d_{x^2-y^2}$-wave superconductors, such as the cuprates~\cite{KosztinLeggett97, Scalapino95}. The relevance of nonlocal effects in an extended temperature regime is controlled by the zero-temperature Ginzburg--Landau parameter, i.e., the ratio of the penetration depth to the coherence length. In Sr$_2$RuO$_4$ this ratio is $\kappa_0 \approx 1.92$~\cite{LandaetaEA23}, placing this material unusually close to the Pippard limit compared to most clean, unconventional superconductors. Finally, we note that the penetration depth was also recently extracted from SQUID susceptometry in thin Sr$_2$RuO$_4$ films, again yielding $\Delta \lambda(T) \sim T^2$ behavior at low $T$~\cite{Ferguson2024}. There, however, the origin of the $T^2$ dependence was interpreted in terms of disorder scattering similar to disordered $d_{x^2-y^2}$-wave superconductors~\cite{HirchfeldGold93}.

Here we perform a theoretical study of the nonlocal electrodynamics specifically relevant for nodal multi-band superconductors. We perform both an exact numerical evaluation of $\Delta \lambda(T)$ and compare this to a generalized nonlocal node-expansion similar to the method by Kosztin and Leggett~\cite{KosztinLeggett97}, but for cases with several distinct nodes in the Brillouin zone, possibly distributed across multiple bands. To examine whether the superconducting order parameter can be constrained from low-temperature penetration depth data, we apply the developed theory to Sr$_2$RuO$_4$. To minimize uncertainties in the description of the normal state, we use a tight-binding model that fits both the experimental Fermi surface and Fermi velocity with unprecedented accuracy. We explore two nodal superconducting gaps with B$_{1\mathrm{g}}$ symmetry, relevant for Sr$_2$RuO$_4$, and compare the node-expansion with the numerically exact result. The analytical analysis shows that the low-temperature slope of $\Delta \lambda (T)$ as a function of $(T/T_c)^2$ is controlled by the sum of reciprocal gap velocities at the order parameter nodes and the corresponding Fermi velocities, the latter are experimentally known from ARPES measurements~\cite{TamaiEA19}. The analysis implies that the penetration depth in isolation is not sensitive to the details of the gap structure since the effect of having one shallow node can be compensated for by instead having two steeper nodes etc. The numerical simulations reveal that both gap structures explored can explain the presently available experimental data for the penetration depth of Sr$_2$RuO$_4$.

%
\section{Penetration depth}
%
Kosztin and Leggett showed that thermal quasiparticles in a clean, type-II, nodal $d$-wave superconductor are responsible for the behavior $\Delta \lambda \sim T^2$ for $T < T^{\ast}$ with $T^{\ast} = \Delta_0 / \kappa_0$, where $\Delta_0$ is gap scale and $\kappa_0 =\lambda_0/\xi_0$ ($\lambda_0$ and $\xi_0$ is the penetration depth and the coherence length, respectively) the zero-temperature Ginzburg--Landau parameter~\cite{KosztinLeggett97}. The key ingredient here are the nonlocal effects (diverging coherence length) experienced by the Cooper pairs formed by momenta near the nodal points, see Fig.~\ref{fig:Geometry}. For such Cooper pairs, it is important to include the spatial variation of the vector potential. \footnote{We note that a different kind of nonlocal Meissner effect based on anisotropies within a multiband London model was studied in Ref.~\cite{Silaev2018}.} This causes the penetration depth increase to acquire an additional factor of $T$ (on top of the $2 k_B T \ln{2} / \Delta_0$ local contribution) coming from the inverse thermal de Broglie wavelength at low temperatures.

In Sr$_2$RuO$_4$ the behaviour $\Delta \lambda \sim T^2$ is observed in a dominant fraction of the temperature window below $T_c \approx 1.5$~K, which is consistent with this material being a marginal type-II superconductor with $\kappa_0 \approx 1.92 > 1/\sqrt{2}$~\cite{LandaetaEA23, BonaldeEA00, MuellerEA23} and $\Delta_0 \approx 0.35$~meV~\cite{MadhavenEA19, MuellerEA23}. Interestingly, Sr$_2$RuO$_4$ is closer to the Pippard limit than most unconventional type-II superconductors. Owing to its clean crystals, this points to the importance of nonlocal electrodynamics to understand its superconducting state, possibly also its response to a weak magnetic field as posed by muons~\cite{HuddartEA21}.

\subsection{Nonlocal electrodynamics}
\begin{figure}[tb]
	\centering
    \includegraphics[width=0.85\linewidth]{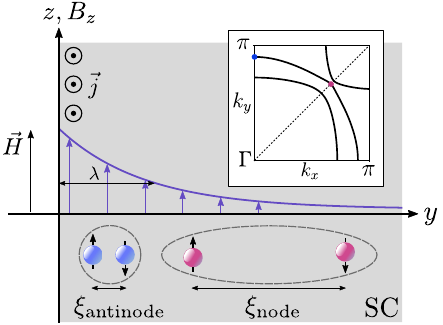}
\caption{Geometry and magnetic field (along $\hat{z}$) penetration into the Meissner state of the superconductor located in the half-plane $y>0$. The magnetic penetration depth is denoted by $\lambda$. Inset: reciprocal space of a tetragonal crystal with the Fermi surface of Sr$_2$RuO$_4$ and the nodal line (dashed) of the B$_{1\mathrm{g}}$ irreducible representation indicated. Close to the nodal point(s), the Cooper pairs are characterized by a long coherence length and hence acquire nonlocal contributions in their electromagnetic response.}
	\label{fig:Geometry}
\end{figure}

In a weak magnetic field, linear response theory for the Meissner state dictates the decay of the magnetic field into the superconductor, as per $j(y) = - \int \D y'~K(y-y') A(y')$. Here, $j$ is the screening supercurrent density (along $\hat{x}$) with the boundary between vacuum and the superconductor at $y=0$, and $A$ is the magnetic vector potential. The geometry is shown in Fig.~\ref{fig:Geometry}.
In dimensionless units, $\tilde K = (4\pi \lambda_0^2/c)K$, and with the kernel difference defined as $\delta \tilde{K}(\tilde{q};T) \equiv \tilde{K}(\tilde{q}; T) - \tilde{K}(\tilde{q};0)$, the penetration depth change can be expressed as
\begin{equation}
\frac{\Delta \lambda(T)}{\lambda_0} = \frac{2}{\pi}\int_0^{\infty} \frac{-\delta \tilde{K}(\tilde{q};T)~\D \tilde{q}}{\big(\tilde{q}^2 + \tilde{K}(\tilde{q};0) \big)\big(\tilde{q}^2 + 1 + \delta \tilde{K}(\tilde{q};T) \big)},
\label{eq:PenetrationDepthMAIN}
\end{equation}
where $\tilde{q} = q\lambda_0$, see Appendix~\ref{sec:Leggett}. We emphasize that this is an exact expression that includes a correction responsible for an upturn in $\Delta \lambda$ at higher $T$ as compared to Ref.~\onlinecite{KosztinLeggett97}. The dimensionless kernel can be computed by means of standard Green's function methods, assuming linear response in the Meissner state and solving the relevant Maxwell equation in the superconductor~\cite{abrikosov2012methods}. In the Matsubara representation the kernel can be expressed as~\cite{abrikosov2012methods, KosztinLeggett97}
\begin{equation}
\tilde{K}(\tilde{q};T) = 2\pi k_B T \!\!\sum_{n=-\infty}^{\infty}\!\! \left\langle \frac{\hat{p}_{\parallel}^2 \Delta_{\bo{p}}^2 }{\sqrt{\omega_n^2 + \Delta_{\bo{p}}^2}(\omega_n^2 + \Delta_{\bo{p}}^2 + \alpha^2)} \right\rangle_{\mathrm{FS}}\!\!, \label{eq:Matsubara}
\end{equation}
where the fermionic Matsubara frequencies are given by $\omega_n = \pi k_B T (2n+1)$, $\hat{p}_{\parallel} = \cos(\theta)$ is the Fermi surface momentum projected onto the boundary ($\theta$ is the polar angle of $\bo{p}$), and the parameter $\alpha = \pi \Delta_0 \tilde{q}  \sin(\theta) /(2\kappa_0)$ is the projected magnetic field penetration and is responsible for the nonlocal effects. Here, we employed the BCS expression for the coherence length, such that $\lambda_0 = v_F \kappa_0 /(\pi \Delta_0)$. Finally, $\Delta_{\bo{p}}$ is the (temperature-dependent) order parameter, and the (dimensionless) Fermi surface average is evaluated as $\langle h \rangle_{\mathrm{FS}} \equiv (\bar{v}_F/\lvert S_F\rvert ) \int_{S_F} \D \hat{\bo{k}}~h(\hat{\bo{k}}) /  v_F(\hat{\bo{k}})$ (see Eq. (\ref{eq:SumDecomposition})) where $v_F(\hat{\bo{k}})$ is the Fermi velocity and $\lvert S_F\rvert$ the Fermi surface area. It can be noted that the zero-temperature kernel, $\tilde{K}(\tilde{q};0)$, has a simple closed-form expression, as stated in Appendix~\ref{sec:Leggett}.

In practice, realistic modelling of $\Delta \lambda/\lambda_0$ in the entire temperature window below $T_c$ is achieved by feeding in a high-precision multiband tight-binding model, as well as experimental values for $\kappa_0$, $\Delta_0$, and $T_c$. Since the sum in Eq.~\eqref{eq:Matsubara} converges rapidly, one can in practice calculate the Fermi surface average for each frequency and truncate the Matsubara series at some $n \gg \frac{\Delta_0}{2\pi k_B T}$ for any non-zero $T$~\footnote{If one instead approaches the problem of evaluating $\delta \tilde{K}$ after recasting the Matsubara sum as an integral, care has to be taken with respect to the integrable singularities that emerge at the endpoints of the $\omega$ integration, cf.~Eq.~\eqref{eq:KernelDiff2}.}. If one models the $T$ dependence of the gap by the interpolation formula $\tanh{(1.74\sqrt{T/T_c - 1})}$, the only remaining degrees of freedom lie in the momentum-dependent gap structure.

\subsection{The node approximation}
Building on the Kosztin--Leggett philosophy at the lowest $T$, useful insights can be harvested by first rewriting Eq.~\eqref{eq:Matsubara} using contour integration techniques, and then linearizing the gap around its nodes, $\Delta_{\bo{p}} \approx \sum_j v_{\Delta,j} (\theta - \theta_j)$, where the thermally active quasiparticles are situated at the lowest $T$. We will henceforth refer to $v_{\Delta,j}$ as the ``gap velocity'' at node $j$. The dimensionless kernel difference, with the intermediate steps shown in Appendix~\ref{sec:Leggett}, can in this case be recast as
\begin{equation}
\begin{aligned}
-\delta \tilde{K}(\tilde{q};T) &\approx k_B T \frac{2\pi  \ln{2}}{\lvert S_F \rvert} \sum_j \frac{2 \hat{p}_{\parallel, j}^2 \bar{v}_F}{v_{F,j} v_{\Delta,j}} \Big\lvert \frac{\partial \bo{k}_F}{\partial \theta} \Big\rvert_j \\
&\hspace{-35pt} \times \left[ 1 - \frac{1}{\ln{2}}\int_{0}^{\alpha_j/T} \D x~\tilde{f}(x) \sqrt{1-\left(x T/\alpha_j \right)^2 } \right],
\label{eq:NodeKernel}
\end{aligned}
\end{equation}
where $\tilde{f}(x) = (1+\exp(x))^{-1}$ is the Fermi function and where the sum runs over distinct nodes in the Brillouin zone (possibly distributed across multiple bands). The symbols are otherwise explained earlier. This result is a multiband generalization of the node approximation proposed by Kosztin and Leggett. To make evaluation fast we further approximate $\tilde{K}(\tilde{q};0) \approx 1 = \tilde{K}(0;0)$ in Eq.~\eqref{eq:PenetrationDepthMAIN} when evaluating the penetration depth from the node approximation in the next section.

To validate the above expression, we benchmark it in the simple $d_{x^2-y^2}$-wave case ($\Delta_{\bo{p}} = \Delta_0 \cos(2\theta)$) using a circular Fermi surface and isotropic Fermi velocity. The four distinct nodes all have $\hat{p}_{\parallel}^2 = 1/2$ and the associated gap velocities are $v_{\Delta} = 2\Delta_0$. In this case we recover the standard result $\delta \tilde{K}(\tilde{q};T) = \delta \tilde{K}(0;T) F(\tilde{q}/t)$, where $t = T/T^{\ast}$, with $T^{\ast} = \Delta_0/\kappa_0$, and $F$ is a universal function similar to the lower line in Eq.~\eqref{eq:NodeKernel}, stated in Appendix~\ref{sec:Leggett}. The prefactor reduces to the well-known $-\delta \tilde{K}(0;T) = 2k_B T \ln{2} / \Delta_0$, i.e., the local result~\cite{HirchfeldGold93, Scalapino95, LiEA00}. The function $F$ ensures that below the characteristic temperature scale $T^{\ast}$, the change in the penetration depth depends quadratically on temperature, $\Delta \lambda \sim T^2$. 

For the multinode generalization in Eq.~\eqref{eq:NodeKernel} the above-mentioned factorization breaks down, and the kernel difference instead depends on all of the distinct $\tilde{q}/t_j$ where $t_j \propto \frac{\Delta_0}{\kappa_0} \sin(\theta_j)$ for each distinct gap node $j$. Thus, each distinct gap node is associated with a characteristic temperature scale, the minimum of which dictates the regime in which $\Delta \lambda \sim T^2$. Some key insights are gained by the details added to Eq.~\eqref{eq:NodeKernel} caused by having a non-isotropic Fermi surface and higher-harmonic gap structure, possibly with multiple distinct nodes. Equation~\eqref{eq:NodeKernel} tells us that $\lim_{T\to T_c} \Delta \lambda(T) / (T/T_c)^2$ is proportional to a weighted sum over the distinct gap nodes, where the weight contains a product of the reciprocal gap velocity and the reciprocal Fermi velocity. The primer is strain-tunable and implies an enhanced sensitivity to gap structures with nodes at strain-induced van Hove points. In the context of Sr$_2$RuO$_4$, no substantial variations in the penetration depth slope change are detected as uniaxial strain is applied~\cite{MuellerEA23}, which argues against both B$_{2\mathrm{g}}$ ($d_{xy}$) and A$_{2\mathrm{g}}$ ($g_{xy(x^2-y^2)}$) type orders, which is also consistent with elastocaloric measurements~\cite{LiEA22}. Generally, however, the dependency on the reciprocal gap velocities implies that penetration depth data alone do not impose any crisp constraints on the momentum structure of the gap, since the effect of having one shallow node can be compensated for by having two steeper nodes.

There are at least two reasons why there are few candidate materials for observing $\Delta \lambda \sim T^2$ from the above mechanism. First, disorder is known to cause a saturation of $\Delta \lambda$ below a characteristic temperature set by the scattering rate~\cite{HirchfeldGold93}, which will effectively blur the observation of a quadratic temperature dependence. Second, the most well-characterized $d$-wave superconductors are strong type-II, $\kappa_0 \gg 1$, suppressing $T^{\ast}$ to a tiny fraction of the gap.
In both of these respects Sr$_2$RuO$_4$ presents a counter-example: there is strong evidence that it is both nodal~\cite{DeguchiEA04, HassingerEA17} and only marginally type-II~\cite{LandaetaEA23}, and it can be prepared to exceptionally high crystal quality~\cite{MackenzieMaeno03}.
Other candidate superconductors possibly relevant to a low-$T$ penetration depth of $\Delta \lambda \sim T^2$ caused by nonlocal electrodynamics include cuprates~\cite{KosztinLeggett97}, KFe$_2$As$_2$~\cite{WilcoxEA22}, and the heavy-fermion material CeCoIn$_5$~\cite{Chia2003, Prozorov_2006,Biderang}. Below we focus on the case of Sr$_2$RuO$_4$.

\section{The case of Strontium Ruthenate}
\label{sec:SRO}
%

%
\subsection{Tight-binding model}
\label{sec:TB}
%
We depart from a standard three-band tight-binding model ansatz for Sr$_2$RuO$_4$,
\begin{equation}
H_0 = \sum_{\bo{k}, \sigma} \vec{\psi}^{\dagger}_{\sigma}(\bo{k}) \pazocal{H}_{\sigma}(\bo{k}) \vec{\psi}_{\sigma}(\bo{k}),
\label{eq:TBHamiltonian}
\end{equation}
where $\vec{\psi}_s(\bo{k}) = [c_{ xz, s}(\bo{k}), \hspace{1mm} c_{yz, s}(\bo{k}), \hspace{1mm} c_{xy, -s}(\bo{k})]^{\mathrm{T}}$, $\sigma = \pm$ denotes spin, and $a\in\{xz,yz,xy\}$ labels the Ru $d$-orbitals (the $t_{2\mathrm{g}}$ triplet), which are the only relevant orbitals close to the Fermi energy~\cite{DamascelliEA14}. 
\begin{figure}[tb]
	\centering
	\includegraphics[width=0.9\linewidth]{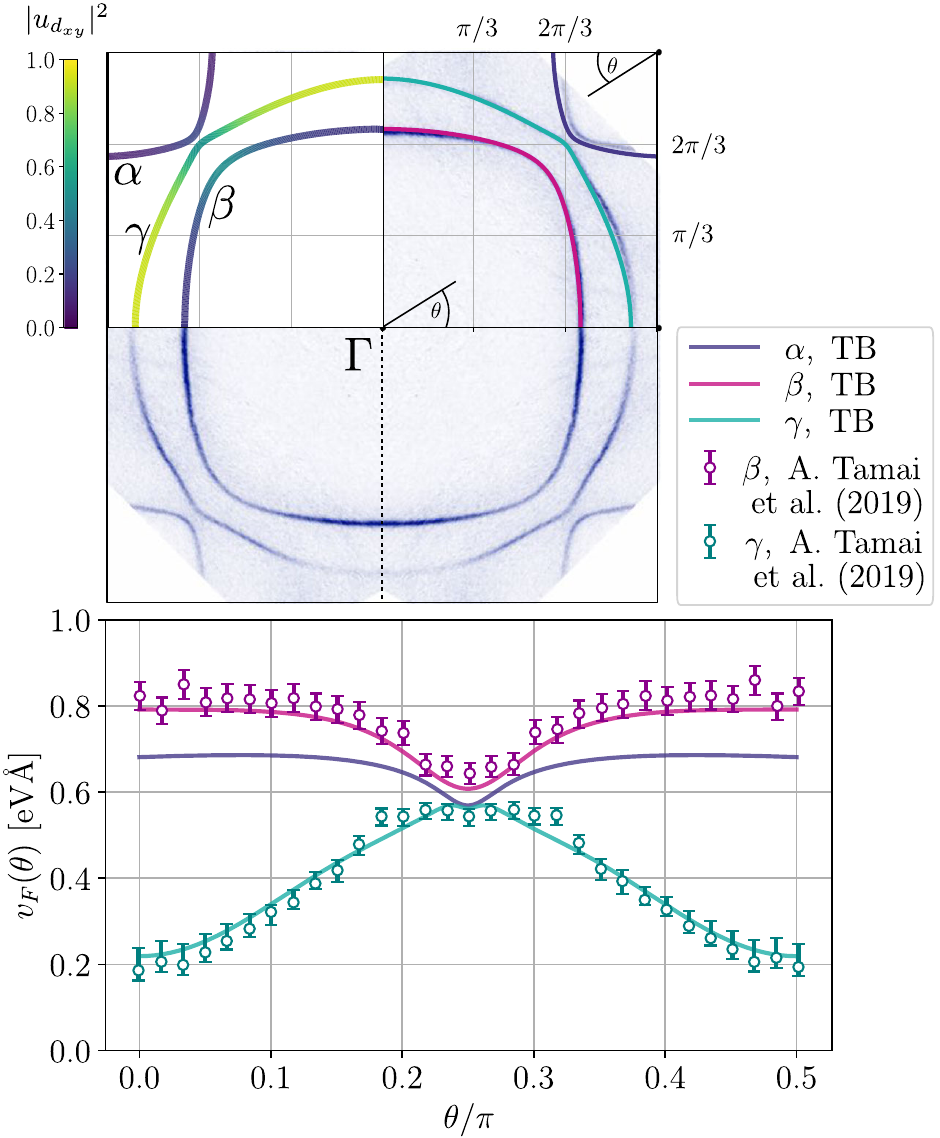}
\caption{Fermi surface and Fermi velocities of Sr$_2$RuO$_4$. Model ($k_z = 0$): solid lines (labelled ``TB'') show the model Fermi surface (top panel) and the Fermi velocities (bottom panel). The second quadrant in the top panel shows the Ru $d_{xy}$ orbital content of the bands. The average model Fermi velocity is $\bar{v}_F = 0.498$~eV\AA. Data: high-resolution ARPES Fermi surface (dark blue, top panel) and the Fermi velocities (circles with error bars, bottom panel) extracted from Ref.~\onlinecite{TamaiEA19}.}
	\label{fig:FS}
\end{figure}
The form of $\pazocal{H}_{\sigma}$ is
\begin{equation}
\pazocal{H}_{\sigma}(\bo{k}) = \begin{pmatrix}
\xi_{xz}(\bo{k}) & \xi_{xz,yz}(\bo{k}) - i \sigma \eta &  i\eta \\
\xi_{xz,yz}(\bo{k})+ i \sigma\eta & \xi_{yz}(\bo{k}) &  - \sigma\eta \\ - i\eta & - \sigma\eta & \xi_{xy}(\bo{k})
\end{pmatrix},
\label{eq:H0}
\end{equation}
where spin-orbit coupling is parametrized by $\eta$ and originates from the (dominant) onsite term $2\eta\sum_i \vec{L}_i \cdot \vec{S}_i$ as projected onto the $t_{2g}$ Ru triplet. Explicit forms of the inter- and intraband energies are listed in Appendix~\ref{sec:AppTB}.

To obtain a quantitatively accurate parametrization, we consider as a starting point the tight-binding parameters from Ref.~\onlinecite{JerzembeckEA21} as derived from relativistic DFT calculations. While these parameters provide a realistic energy scale and Fermi surface, they still do not quantitatively match the Fermi velocities as extracted from high-resolution ARPES measurements for bands $\beta$ and $\gamma$ in Ref.~\onlinecite{TamaiEA19}. To correct for this discrepancy, which is also present in another tight-binding model widely employed in the literature~\cite{ZabolotnyyEA13}, we manually tune tight-binding parameters until a reasonable match with both the experimental Fermi surface \emph{and} Fermi velocity $v_F$ is obtained. The result is shown in Fig.~\ref{fig:FS}. The effective model, with the set of parameters listed in Appendix~\ref{sec:AppTB}, provides a high-precision effective normal state description of Sr$_2$RuO$_4$. We stress that a model that is qualitatively and quantitatively accurate in the above respect could be crucial to accurately perform calculations sensitive to $v_F$.

%
\subsection{Numerical evaluation of the penetration depth}
\label{sec:Numerics}
%
Equipped with an accurate description of the normal state, we now turn to the evaluation of the penetration depth difference, $\Delta \lambda(T)/\lambda_0$, for some illustrative order parameters. 

\begin{figure}[tb]
	\centering
	\includegraphics[width=\linewidth]{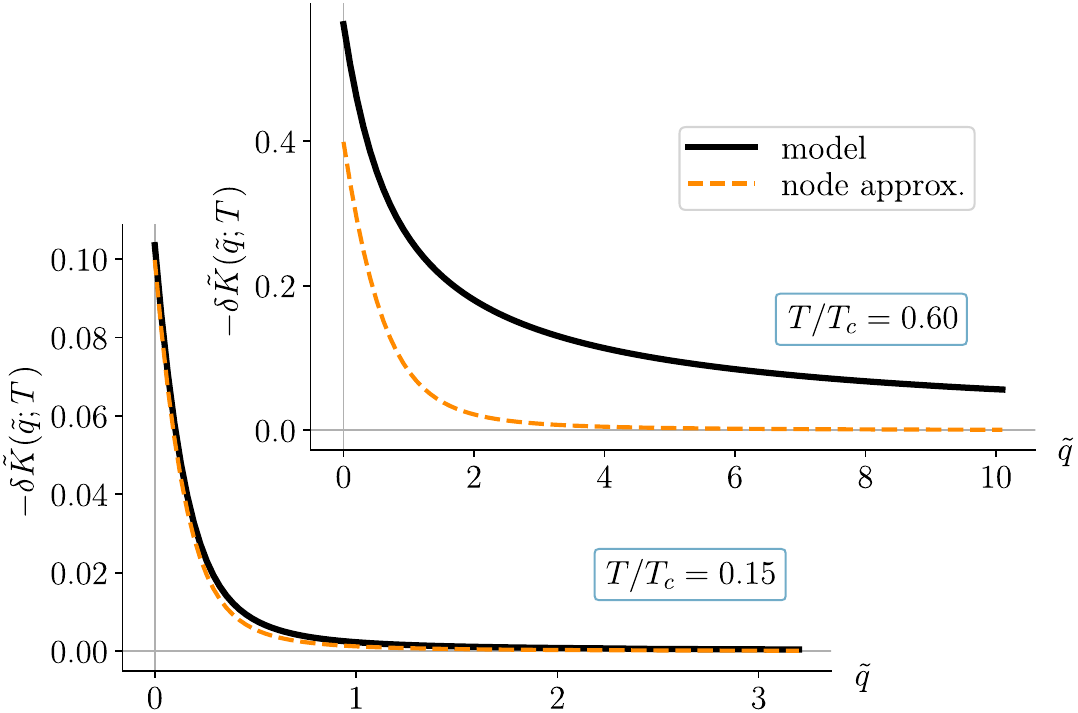}
\caption{Dimensionless kernel differences, $-\delta \tilde{K}(\tilde{q};T)$, calculated at $T = 0.15T_c$ and $T = 0.6T_c$ for both the node approximation of Eq.~\eqref{eq:NodeKernel} (``node approx.'') and in the Matsubara representation (``model''). We use the three-band tight-binding model of Sec.~\ref{sec:TB} and order parameters of Fig.~\ref{fig:PenetrationDepth}(a). Due to the curvature of $\Delta(\theta)$, the node approximation overestimates the gap and underestimates the penetration depth at higher temperatures.}
	\label{fig:KernelDiff}
\end{figure}
\begin{figure*}[b!th]
	\centering
	\includegraphics[width=\linewidth]{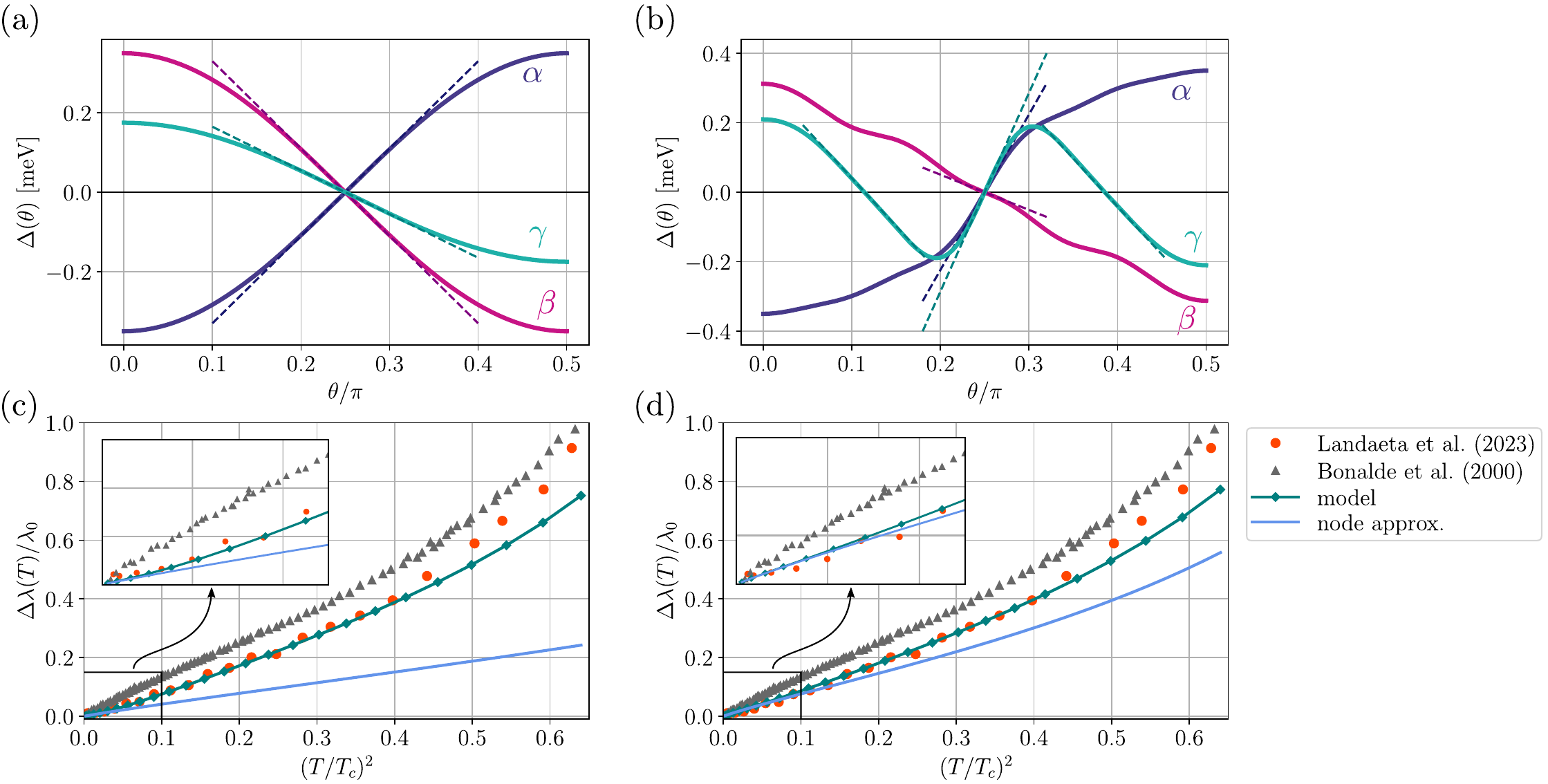}
\caption{Representative B$_{1\mathrm{g}}$ order parameters [(a) and (b)], and associated penetration depths [(c) associated with (a), and (d) associated with (b)] using both the node approximation (``node approx.'') and the Mastubara representation (``model'') to calculate the kernel. We use the three-band tight-binding model of Sec.~\ref{sec:TB}. Experimental data points from Refs.~\onlinecite{BonaldeEA00, LandaetaEA23} are plotted as gray triangles and orange disks for comparison. In (a) and (b) linearizations around the gap nodes are shown with dashed lines.}
	\label{fig:PenetrationDepth}
\end{figure*}

First, we calculate the kernel difference $-\delta \tilde{K}(\tilde{q};T)$ at two temperatures using both the node approximation of Eq.~\eqref{eq:NodeKernel} and the full Matsubara sum of Eq.~\eqref{eq:Matsubara}. For illustration we employ the simplest gap structure consistent with B$_{1\mathrm{g}}$ symmetry, i.e., $\Delta_0 \cos(2\theta)$ on all bands with $\Delta_0 = 0.35$~meV on bands $\alpha$ and $\beta$, and half the magnitude on band $\gamma$. These values are motivated by STM experiments~\cite{MadhavenEA19, MuellerEA23}, and the gap magnitude on $\gamma$ was reduced to match the experimental penetration depth slope at low $T$. We otherwise fixed $\kappa_0 = 1.92$ and $T_c = 1.5$~K, in agreement with experiments~\cite{LandaetaEA23}. The resulting kernel differences  $-\delta \tilde{K}(\tilde{q};T)$ are shown in Fig.~\ref{fig:KernelDiff}, and the order parameter and penetration depth are shown in Fig.~\ref{fig:PenetrationDepth}(a) and (c).

Comparing the penetration depth calculations with the experimental data reveals that the temperature window in which the node approximation matches with the realistic modelling is roughly $T/T_c \lesssim 0.2$. The primary reason for this is that the node approximation based on the slope $v_\Delta$ overestimates the value of the order parameter $\lvert \Delta(\theta) \rvert$ when linearizing the gap. For a given temperature $T$, there are fewer quasiparticle states available in the node approximation compared to the full dependence of $\Delta(\theta)$, causing the node approximation to, in this case, monotonically underestimate the penetration depth. Additionally, the $T$ dependence of the gap and the correction posed by the denominator of Eq.~\eqref{eq:PenetrationDepthMAIN} both contribute with an upturn in $\Delta \lambda(T)$ close to the transition temperature $T_c$, approximately consistent with the experimental data. The realistic modelling shows that a simple $d_{x^2-y^2}$-wave gap is sufficient to explain the experimental data, albeit with a slight discrepancy at the highest $T$.

In Fig.~\ref{fig:PenetrationDepth} (b) and (d) we show the results of calculating the penetration depth for a more involved order parameter, still within the B$_{1\mathrm{g}}$ irreducible representation, but with an angular dependence inspired by weak-coupling and RPA spin-fluctuation calculations~\cite{RoisingEA19,RomerEA19,RomerMaier2020,ShengEA22}. In this case, the node approximation performs better when comparing to the realistic modelling. Since the order parameter is more involved by having contributions from multiple harmonics, the error introduced by linearizing the gap is non-monotonic, i.e., the gap is both over- and underestimated on the various bands. As the realistic  modelling shows, an equally convincing $T$-dependent penetration depth is obtained for this order parameter. Therefore, not surprisingly, the change in the penetration depth is not capable of resolving subtle differences in the the nodal structure of the order parameter. As the nodal expansion reveals, the slope of $\Delta \lambda (T)$ as a function of $(T/T_c)^2$ at the lowest temperatures is a weighted sum of reciprocal gap velocities at the nodes, so the increase in slope gained by an additional node can be compensated for by increasing the gap velocity of one or more nodes. 

In the above analysis we have focused on $d$-wave ordering since it appears as a strong contender for the leading superconducting order relevant for Sr$_2$RuO$_4$. We note, however, that purely based on the penetration depth data, only the nodal structure of the gap function matters. Thus, one may also obtain equally good fits to the experimental data shown in Fig.~\ref{fig:PenetrationDepth} (c,d) from, for example, a nodal $s$-wave order parameter. Finally, fully gapped order parameters such as the (earlier) proposed triplet state of type $p_x+ip_y$ is incapable of explaining the penetration depth data of Sr$_2$RuO$_4$.

\section{Conclusions}
In summary, we have calculated the temperature-dependent change of the penetration depth $\Delta \lambda$ including nonlocal effects from line nodes. We have provided both a multi-band multi-node generalization of the Kosztin-Leggett result~\cite{KosztinLeggett97}, and demonstrated a straightforward numerical procedure to evaluate $\Delta \lambda$ numerically exact, given a multi-band tight-binding description. Focusing on the case of Sr$_2$RuO$_4$, posing as a prime candidate material owing to its clean crystals and evidence of nodal order, we investigated two $d$-wave gap structures with different nodal properties. The analysis reveals that the low-temperature penetration depth is sensitive to the sum of reciprocal gap velocities at the nodes, and that both orders investigated have nodal properties compatible with the presently available data.

\section{Acknowledgements}

We acknowledge useful discussions with C.~Hicks.
H.S.R.~was supported by research Grant No.~40509 from VILLUM FONDEN. A.K.~acknowledges support by the Danish National Committee for Research Infrastructure (NUFI) through the ESS-Lighthouse Q-MAT.

\appendix

%
\section{Generalized Kosztin--Leggett theory}
\label{sec:Leggett}
%
Here, we re-derive and generalize the central result of Kosztin and Leggett~\cite{KosztinLeggett97} for the penetration depth of a nodal type-II superconductor.

In a weak magnetic field, the Meissner state of a superconductor responds linearly to the perturbation, 
\begin{equation}
j(y) = - \int \D y' K(y-y') A(y').
\label{eq:CurrentDensity} 
\end{equation}
Here, $j$ is the screening supercurrent density (pointing along $\hat{x}$, $y = 0$ is the position of the superconductor boundary), $K$ is the electromagnetic response kernel, and $A$ is the magnetic vector potential. With a specular boundary, the magnetic penetration depth is given by
\begin{equation}
\frac{\lambda(T)}{\lambda_0} = \frac{2}{\pi} \int_0^{\infty} \frac{\D \tilde{q}}{\tilde{q}^2 + \tilde{K}(\tilde{q};T)},
\label{eq:PenetrationDepth}
\end{equation}
where $\tilde{q} = \lambda_0 q$ with $\lambda_0 \equiv \lambda(0)$, and $\tilde K = (4\pi \lambda_0^2/c)K$. The dimensionless kernel satisfies $\tilde{K}(\tilde{q}\to 0; 0) = 1$ in the local limit. Writing $\tilde{K}(\tilde{q};T) = 1  + \delta \tilde{K}(\tilde{q};T)$, with $\delta \tilde{K}(\tilde{q};T) \equiv \tilde{K}(\tilde{q}; T) - \tilde{K}(\tilde{q};0)$ leads to the exact expression for $\Delta \lambda(T) \equiv \lambda(T) - \lambda_0$ given in the main text Eq.~\eqref{eq:PenetrationDepthMAIN}. The zero temperature kernel can be evaluated analytically~\cite{KosztinLeggett97}, with 
\begin{equation}
\tilde{K}(\tilde{q};0) = 1 - \left\langle 2\hat{p}_{\parallel}^2 \left[ 1 - \frac{\mathrm{arcsinh}(\alpha/\Delta_{\bo{p}})}{\alpha/\Delta_{\bo{p}} \sqrt{1 + (\alpha/\Delta_{\bo{p}})^2} } \right] \right\rangle_{\mathrm{FS}}.
\label{eq:MatsubaraZeroT}
\end{equation}

Using contour integration techniques~\cite{abrikosov2012methods}, the response kernel correction can be evaluated as
\begin{equation}
\begin{aligned}
-\delta\tilde{K}(\tilde{q};T) &= 2 \int_0^{\infty} \D\omega~f(\omega) \\
&\times \left\langle 2\hat{p}_{\parallel}^2 \mathrm{Re} \frac{\Delta_{\bo{p}}^2}{\sqrt{\omega^2-\Delta_{\bo{p}}^2}(\Delta_{\bo{p}}^2-\omega^2+\alpha^2)} \right\rangle_{\mathrm{FS}},
\end{aligned}
\label{eq:KernelDiff}
\end{equation}
where $f(\omega) = (1+\exp(\beta \omega))^{-1}$ is the Fermi function, $\Delta_{\bo{p}}$ is the order parameter, $\hat{p}_{\parallel}  = \cos(\theta)$ is the projection of the Fermi surface momentum on the $\hat{x}$-axis, and $\alpha = q v_F \sin{\theta} /2$. Setting $\alpha = 0$ reproduces the local result~\cite{HirchfeldGold93}.

To evaluate the Fermi surface average, we first recast momentum sums as integrals over $(\xi,\hspace{1mm} \hat{\bo{k}})$, where $\hat{\bo{k}}$ lies on the Fermi surface defined by $S_F(\xi) \equiv \lbrace \hat{\bo{k}} :~\xi_{\hat{\bo{k}}} = \xi \rbrace$ in the following manner (introducing also the electronic cutoff $\omega_c$):
\begin{equation}
\sum_{\bo{k}:~\lvert \xi_{\bo{k}} \rvert < \omega_c} h_{\bo{k}} = \int_{-\omega_c}^{\omega_c} \D\xi \hspace{1mm} \rho_{\xi} \int_{S_F(\xi)} \f{\D \hat{\bo{k}}}{\lvert S_F\rvert } \f{\bar{v}_F}{v_F(\hat{\bo{k}})} h(\hat{\bo{k}}),
\label{eq:SumDecomposition}
\end{equation}
where the Fermi velocity, the average Fermi velocity, and the density of states are given by
\begin{align}
v_F(\hat{\bo{k}}) &= \lvert \nabla \xi_{\hat{\bo{k}}} \rvert, \label{eq:FermiVelocity} \\
\bar{v}_F &= \Big[ \int_{S_{F}(\xi)} \frac{\D\hat{\bo{k}}}{\lvert S_{F} \rvert} \f{1}{v_F(\hat{\bo{k}})} \Big]^{-1}, \label{eq:SurfaceArea} \\
\rho_{\xi} &= \int_{S_F(\xi)} \f{\D \hat{\bo{k}}}{(2\pi)^d} \f{1}{v_F(\hat{\bo{k}})}, \label{eq:DensityOfStates}
\end{align}
respectively, and where $\lvert S_F \rvert$ is the Fermi surface area. From the above we define the (dimensionless) Fermi surface average as
\begin{equation}
	\langle A \rangle_{\mathrm{FS}} \equiv  \int_{S_F(\xi)} \f{\D \hat{\bo{k}}}{\lvert S_F\rvert } \f{\bar{v}_F}{v_F(\hat{\bo{k}})} A,
\label{eq:FSaverage}
\end{equation}
such that $\langle 1 \rangle_{\mathrm{FS}} = 1$. 

We next expand the order parameter around its nodes, situated at angles $\theta_j$, $\Delta_{\bo{p}} \approx \sum_j v_{\Delta,j} (\theta - \theta_j)$, where $v_{\Delta,j}$ is the ``gap velocity'' of node $j$. Close to the nodes (at low temperatures), we can safely ignore the angular dependence of $\alpha$ and $\hat{p}_{\parallel}$. Since Eq.~\eqref{eq:KernelDiff} picks up contributions around each node, we then get
\begin{widetext}
\begin{equation}
\begin{aligned}
-\delta\tilde{K}(\tilde{q};T) &= 2 \int_0^{\infty} \D\omega~f(\omega) \left\langle 2\hat{p}_{\parallel}^2 \mathrm{Re} \frac{\Delta_{\bo{p}}^2}{\sqrt{\omega^2-\Delta_{\bo{p}}^2}(\Delta_{\bo{p}}^2-\omega^2+\alpha^2)} \right\rangle_{\mathrm{FS}} \\
&\approx \frac{2}{\lvert S_F \rvert} \int_0^{\infty} \D\omega~f(\omega) \sum_j \frac{2\hat{p}_{\parallel, j}^2  \bar{v}_F }{v_{F,j}v_{\Delta,j}} \Big\lvert \frac{\partial \bo{k}_F}{\partial \theta} \Big\rvert_j \int_{-\omega}^\omega \mathrm{Re} \frac{\D u~u^2}{\sqrt{\omega^2 - u^2}(u^2 -\omega^2 + \alpha_j^2)} \\
&= \frac{2\pi}{\lvert S_F \rvert} \int_0^{\infty} \D\omega~f(\omega) \sum_j \frac{2\hat{p}_{\parallel, j}^2 \bar{v}_F}{v_{F,j} v_{\Delta,j}} \Big\lvert \frac{\partial \bo{k}_F}{\partial \theta} \Big\rvert_j \mathrm{Re}\left[ 1 - \sqrt{1-\omega^2/\alpha_j^2} \right] \\
&= k_B T \frac{2\pi  \ln{2}}{\lvert S_F \rvert} \sum_j \frac{2 \hat{p}_{\parallel, j}^2 \bar{v}_F}{v_{F,j} v_{\Delta,j}} \Big\lvert \frac{\partial \bo{k}_F}{\partial \theta} \Big\rvert_j \left[ 1 - \frac{1}{\ln{2}}\int_{0}^{\alpha_j/T} \D x~\tilde{f}(x) \sqrt{1-\left(x T/\alpha_j \right)^2 } \right],
\end{aligned}
\label{eq:KernelDiff2}
\end{equation}
%
where $\alpha_j = \tilde{q} v_{F,j} \sin(\theta_j) / (2\lambda_0 )$, $\tilde{f}(x) \equiv (1+\exp(x))^{-1}$ is the Fermi function with dimensionless argument, and where the sum runs over distinct nodes in the Brillouin zone (possibly distributed across multiple bands). This result is a multiband generalization of the node approximation proposed by Kosztin and Leggett. This low-temperature approximation is valid in the temperature window in which the order parameter can be reasonably approximated by a linear function of the angle deviation from the node.

To validate the generalized (multiband) expression of Eq.~\eqref{eq:KernelDiff2}, we evaluate it for the simple $d$-wave order parameter $\Delta_{\bo{p}} = \Delta_0 \cos(2\theta)$ and a circular Fermi surface with isotropic Fermi velocity~\cite{KosztinLeggett97}. The four distinct nodes all have $\hat{p}_{\parallel}^2 = 1/2$ and gap velocity $v_{\Delta} = 2\Delta_0$. Further writing $\lambda_0 = \kappa_0 \xi_0$, and using the BCS expression for the coherence length, $\xi_0 = v_F /(\pi \Delta_0)$, leads to
\begin{equation}
\begin{aligned}
-\delta\tilde{K}(\tilde{q}; T) &= \frac{2k_B T \ln{2}}{\Delta_0}
\left[ 1 - \frac{1}{\ln{2}} \int_0^{\pi \frac{\sqrt{2} z}{4}} \D x~\tilde{f}(x)\sqrt{1 - \frac{8 x^2}{\pi^2 z^2} } \right],
\end{aligned}
\label{eq:KernelDiff3}
\end{equation}
where $z \equiv \frac{\tilde{q}}{T/T^\ast}$, and $T^{\ast} \equiv \frac{\Delta_0}{\kappa_0}$. Here $\kappa_0$ is the zero-temperature Ginzburg--Landau parameter, and the prefactor is recognized as $-\delta \tilde{K}(0;T) = 2k_B T \ln{2} / \Delta_0$.

%
\section{Details of the tight-binding model}
\label{sec:AppTB}
%

The inter- and intra-orbital energies in Eqs.~\eqref{eq:TBHamiltonian} and \eqref{eq:H0} take the form
%
\begin{align}
\xi_{xz}(\bo{k}) &= - t_1 \cos{k_x} - t_2 \cos{k_y} - t_3 \cos{k_x} \cos{k_y} - t_4\cos(2k_x) - t_5 \cos(2k_x) \cos{k_y} \nonumber \\ 
&\hspace{20pt} - t_6 \cos(3k_x) - t_7 \cos(k_x/2)\cos(k_y/2)\cos(k_z/2) - \mu_1, \label{eq:xzhopping} \\
\xi_{yz}(k_x, k_y, k_z) &= \xi_{xz}(k_y, k_x, k_z), \label{eq:yzhopping} \\
\xi_{xy}(\bo{k}) &= - t_8 \left[ \cos{k_x} + \cos{k_y} \right] - t_9 \left[ \cos(2k_x) + \cos(2k_y) \right] - t_{10} \cos{k_x} \cos{k_y} \nonumber \\
&\hspace{20pt} - t_{11} \left[ \cos{k_x}\cos(2k_y) + \cos(2k_x)\cos{k_y} \right] - t_{12} \cos(2k_x)\cos(2k_y) \nonumber \\
&\hspace{20pt} - t_{13} \left[ \cos{k_x}\cos(3k_y) + \cos(3k_x)\cos{k_y} \right] - t_{14} \left[ \cos(3k_x) + \cos(3k_y) \right]  \nonumber \\ 
&\hspace{20pt}  - t_{15}\cos(k_x/2)\cos(k_y/2)\cos(k_z/2) - \mu_2 , \label{eq:xyhopping} \\
\xi_{xz,yz}(\bo{k}) &= - t_{16} \sin(k_x/2) \sin(k_y/2) \cos(k_z/2). \label{eq:xzxyhyb}
\end{align}
Tight-binding parameters providing a fit to both the Fermi surface \emph{and} the Fermi velocity of the data in Ref.~\onlinecite{TamaiEA19} are listed in Tab.~\ref{tab:HoppingParameters1} and \ref{tab:HoppingParameters2}. These parameters are largely taken from the relativistic DFT calculation of Ref.~\onlinecite{JerzembeckEA21}. In particular, the $k_z$ dependent terms, responsible for the out-of-plane warping, are identical. The calculations presented in the main text were done with the effective 2D model obtained by fixing $k_z = 0$.

\begin{table}
\centering
\caption{Tight-binding parameters for Eqs.~\eqref{eq:H0}, \eqref{eq:xzhopping}, \eqref{eq:yzhopping} consistent with high-resolution ARPES measurements~\cite{TamaiEA19}.}
\begin{tabular}{p{2.2cm} p{1.2cm} p{1.2cm} p{1.2cm} p{1.2cm} p{1.2cm} p{1.2cm} p{1.2cm} p{1.2cm} p{1.2cm}}  
\toprule
Parameter & $t_1$ & $t_2$ & $t_3$ & $t_4$ & $t_5$ & $t_6$ & $t_7$ & $\mu_1$ & $\eta$ \\ \hline
Value [meV] & $562.7$ & $99.9$ & $-47.3$ & $-174.3$ & $-51.8$ & $-11.0$ & $102.0$ & $209.9$ & $81.0$ \\ \bottomrule
\label{tab:HoppingParameters1}
\end{tabular}
\end{table}

\begin{table}
\centering
\caption{Tight-binding parameters for Eqs.~\eqref{eq:xyhopping} and \eqref{eq:xzxyhyb} consistent with high-resolution ARPES measurements~\cite{TamaiEA19}.}
\begin{tabular}{p{2.2cm} p{1.2cm} p{1.2cm} p{1.2cm} p{1.2cm} p{1.2cm} p{1.2cm} p{1.2cm} p{1.2cm} p{1.2cm} p{1.2cm}}  
\toprule
Parameter & $t_{8}$ & $t_9$ & $t_{10}$ & $t_{11}$ & $t_{12}$ & $t_{13}$ & $t_{14}$ & $t_{15}$ & $\mu_2$ & $t_{16}$ \\ \hline
Value [meV] & $458.2$ & $-7.5$ & $330.0$ & $25.3$ & $32.8$ & $8.8$ & $3.5$ & $-12.3$ & $284.2$ & $72.4$  \\ \bottomrule
\label{tab:HoppingParameters2}
\end{tabular}
\end{table}
\end{widetext}

\bibliography{Refs}

\end{document}